\documentstyle[12pt]{article}
\begin{document}

\begin{center}

{\Huge Phase Transition in Random

\vspace{0.3 cm}

Networks with Multiple States}

\vspace{0.5 in}

{\Large
 Ricard V. Sol\'e $^{1,2}$, Bartolo Luque $^{3}$ and Stuart Kauffman $^2$}

\vspace{0.5 in}

(1) Complex Systems Research Group, Department of Physics - FEN

Universitat Polit\'ecnica de Catalunya

Campus Nord, M\`odul B5 , 08034 Barcelona (Spain)

\vspace{0.25 in}

(2) Santa Fe Institute

1399 Hyde Park Road, Santa Fe, New Mexico (USA)

\vspace{0.25 in}
(3) Centro de Astrobiolog\'{\i}a (CAB), Ciencias del Espacio, INTA 

Carretera de Ajalvir km. 4, 28850 Torrej\'on de Ardoz, Madrid (Spain)

\end{center}

\baselineskip=5 mm

\vspace{0.6 in}

\begin{abstract}

The critical boundaries separating ordered from chaotic behavior in randomly
 wired $S$-state networks are calculated. These networks are a natural
 generalization of random Boolean nets and are proposed as on extended approach
 to genetic regulatory systems, sets of cells in different states or collectives
 of agents engaged into a set of $S$ possible tasks. A order parameter
 for the transition is computed and analysed. The relevance of these networks to biology, their
 relationships with standard cellular automata and possible extensions are outlined.

\end{abstract}

\begin{center}

\vspace{2.5 cm}

{\large \bf Submitted to Physical Review E}

\end{center}

\newpage

\section{Introduction}

Cellular automata (CA) [1] and random Boolean networks (RBN) [2] are two key
approaches in our understanding of complexity [3,4]. In both cases, discrete-time and
 discrete-space are used, together with a finite number
of states. For CA, we have a dynamical evolution defined
(for the one-dimensional case with a $(2r+1)$-neighborhood) by means of:
$$ \sigma_i(t+1) = \Phi \Bigl [ \;
\sigma_{i-r} (t) , \sigma_{i-r+1}(t) , ... , \sigma_{i+r}(t) \; \Bigr ] \eqno (1) $$
 where $i=1,...,N$ and $\sigma_i(t) \in \Sigma \equiv \{ 0, 1, ... , S-1 \}$. Here
 $\Phi$ is a given
(previously defined) rule [1]. Identical rules are defined for each
$(2r+1)$-neighborhood. In contrast with this homogenity in neighbors and
 rules, RBN's are a particularly interesting class of 
dynamical systems which shares some properties with CA models, but
where randomness is introduced at several levels. Now the dynamics is
given by:
$$ \sigma_i(t+1) = \Lambda_i \; \Bigl [ \;
\sigma_{i_1}(t) , \sigma_{i_2}(t) , ... , \sigma_{i_K}(t) \; 
\Bigr ] \eqno (2) $$
where $i=1,...,N$ and $\sigma_i(t) \in \Sigma \equiv \{0,1\}$. Each automaton is
randomly connected with exactly $K$ others wich send inputs to him. Here
$\Lambda_i$ is a Boolean function also randomly choosen from a set
${\cal F}_K $ of all possible $K$-inputs Boolean functions.
In spite of the random selection both of neighbors and
Boolean functions, it is well known that a critical connectivity $K_c$
exists where "spontaneous order crystallizes" [2].  In $K_c$ a
small number of attractors $(\approx O(\sqrt N))$ is observed which show
high homeostatic stability (i.e. high stability against minimal
perturbations in single elements or Boolean functions), and
low reachability among different attractors. These properties are
consistent with some observations of the genome organization [2,5]. A considerable effort has been dedicated over the last years to the analysis (both theoretical and computational) of the scaling behavior of the numbers and size of cycles close to the critical boundary [5-7].

Alternative models to this problem are provided by neural network-like aproaches [8]. In this case we have a dynamical system given by the set:
$$ \sigma_i(t+1) = \Phi \lbrack J_{ii}\sigma_i(t)+ \sum_{l=1}^{K-1} J_{ij_{l}(i)}\sigma_{j_{l}(i)}(t)+\theta_i\rbrack \eqno (3)$$ where $\Phi(x)$ is a sigmoidal function which can be assimptotically approched by a sign function, $\Phi(x)= sgn (x)$ which is $sgn(x) =0$ if $x\leq 0$ and $sgn (x) = 1$ otherwise. Not surpringly, neural networks with random connectivities show some of the properties displayed by RBN. Two well-defined phases are also described, which also depend upon the input connectivity [9].

Though the picture of on-off genes is a simplified one, in some cases it can be justified
from general arguments based on mutual inhibition among genes [10]. Let
us consider two given genes which interact among them by means of two given
proteins, whose concentrations are indicated by the pair $(x,y)$. This can be
 used as a toy model of some well known properties of the life cycle of
 viruses (like the $\lambda$-Phage [3,10]). A theoretical
model of mutual inhibition is given by the following couple of cross-inhibitory equations:
$${dx \over dt} = {\mu^n \over \mu^n + y^n} - x \eqno (4.a)$$
$${dy \over dt} = {\mu^n \over \mu^n + x^n} - y \eqno (4.b)$$
where $x,y \in \bf R^+$. Here $n \in \bf N$ is a measure of the strength of the
interaction. For $\mu=1/2$, the stability of the fixed point $P^*=(1/2,1/2)$
is given by the eigenvalues of the Jacobi matrix:
$$ L_{\mu}(P^*) = \left ( \matrix{ 
        -1 &  {-{n \over 2}} \cr
         {-{n \over 2}} & -1  \cr} \right ) \eqno (5)$$
which leads to stability for $n<2$ and to instability otherwise. For
$n>n_c=2$, $P^*$ becomes a saddle point and two new stable attractors
$\{ P_+, P_-\}$ are formed (stable nodes). For strong interactions, one of the
concentrations dominate over the other and, basically, the two new attractors
are simply $P_+ \approx (1,0)$ and $P_- \approx (0,1)$. So the result of
this interaction between continuous quantities leads to a basically binary
outcome [11]. 

 But gene activity can lead to more complex
 combinations (see figure (1) as example). And in fact experimental analysis of RNA levels in developing embryos shows a wide range of gene activity levels. Different genes interact in very complex ways, and activation or represion takes place in such a way that the same gene can be more or less active depending on the coupling of proteins to different regulatory sites [12]. To see this, let us consider a general network formed
by $m$ elements showing mutual inhibition, i.e.: 
$${dx_i \over dt} = f_{\mu}^{(i)} (x_1, x_2, ... , x_n)
 = {\mu^n \over \mu^n + \sum_{j \ne i} x_j^n} - x_i \eqno (6)$$
With $i=1,...,m$.
This network leads to the emergence of a (usually large) number of attractors when $n$
is large enough. An example of the phase space trajectories for $m=3$ 
and $\mu=1/2$ is shown in figure 2. We can see in figure (2) upper that for low $n$ values ($n=1$) only a coexistence point is reached. However, see figure (2) lower, once the nonlinearity is increased beyond a treshold, different attractors are obtained,
indicated by the flow of the vector field (here $n=4$). Three corners are
 occupied by the simplest attractors (molecular switches with $P^* \in \{(1,0,0),
 (0,1,0), (0,0,1)\}$). But we also have $P^*=(x_1^*, x_2^*, x_3^*)$
with $x_1^* \neq x_2^* \neq x_3^* \in \{ \alpha, \beta, \gamma \}$ (here
 $\alpha=0.13, \beta=0.34, 
\gamma = 0.97$). So  three different states are available for each 
variable. For a more complex network, many possible states will be possible
 displaying a range of intermediate states. In order to avoid this problem
other approximations have been proposed. These approximations retain the discreteness 
of the units but consider a continuous
 range of values [11, 14] more close to the real picture, where gene activity can
 be measured in terms of RNA concentrations [12]. 

These results makes necessary a generalization of RBNs
extending them into a more general framework. In particular, we can
ask how are the critical boundaries defined 
for a more general situation when multiple states are available.
 
Although the starting point of our study deals whith genomic regulatory circuits,
 a set of different, but related problems could be explored within this framework.
 If a given state is identified in terms of a cell type, for example, the interactions
 between differents cells (automata) with different cell type will eventually lead
 to transitions which are known to be rather complex and neighbour-dependent [12].
 Another example could be a system where each unit can have a complex internal state
 roughly characterized by scalar quantity (or "state") which can be, for example,
 the mean activity. A further scenario where our description applies is a system
 of machines (computers or agents) each one engaged in a given task. These machines
 might be connected in complex ways, and switch from a task to another by depending
 on their actual inputs. 

Discrete dynamical systems involving a given range of states per unit are well known
 in the physics literature. Cellular automata (as described by equation (1)) are
 just the best known example [1]. Another example from statistical mechanics is the
 $q-$ Potts model [13] which generalizes the classical Ising spin systems.
 It well known that this model shows phase transitions with a critical temperature
 $T_c(q)$ depending on $q$, the number of possible spin states.

Our goal in this paper is to explore the properties of the critical points
 for $S$-state random networks. In this context, the analysis of transition 
phenomena in CA shows that, for large r-neighborhoods, a sharp (second order )
 transition occurs for well-defined $\lambda$-values. However this result cannot
 be traslated to the genome because typically gene regulation is a non-local
 phenomenon [8,12]. We should expect, however, to find some agreement between the
 CA results and those derived from networks involving
 multiple states and random connectivity.

The paper is organized the follows manner: in section 2, random networks with multiple states are define and 
the critical surface
 in the $(K,S,p)$ space is derived from Derrida's annealed approximation.
 In the the Appendix I associated with this section a maximum entropy variational approach is used in order to compute 
the expected distribution of states at criticality. In section 3 the relationship with CA models for large number of states $S$ is analysed.
 In section 4 a variant 
of Flyvbjerg's method of stable core computation is introduced in order
 to obtain an order parameter (the {\em self-overlap}) for the transition. In 
the Appendix II associated with this section we compute a Lyapunov exponent for the system. Finally, a discussion of the general results and possible extensions is given in section 5.

\section {The model and Derrida's annealed approach}

 The previous definition of random Boolean network, where
$\Sigma = \{ 0,1 \}$ can be generalized to a wider set of states,
$\Sigma = \{ 0,1, ... , S-1 \}$ that is to say, to random networks
with multiple states (RNS). As with RBN, each element $\sigma_i$
$(i=1,...,N)$ receives $K$ inputs;
so we have:
$$ \sigma_i(t+1) = \Lambda_i^s \; \Bigl [ \;
\sigma_{i_1} (t), \sigma_{i_2}(t) , ... , \sigma_{i_{K}}(t) \; \Bigr ] \eqno (7) $$
where the superindex $s$ of the functions $\Lambda_i^s $ indicate that 
its inputs (and outputs) can take $S$ values and the functions $\Lambda_i^s$
are randomly chosen from a set $ \cal S_K$.

 As with RBNs, this model
exhibits three characteristic dynamical regimes, which we will explore.
Three examples are shown in figures (3-5) for a $N=100$ network with
$S=5$ states, $K=2$ and different $p$-values. Here $p$ is an additional
which was introduced by B. Derrida et al. [15] in order to make the RBN
 transition smooth and continuous. The parameter $p$, known as the bias of
 $\Lambda_i$, is defined as the probability of having $0$ as output in the
function $\Lambda_i$. This parameter is easy to extend in natural form for 
the RNS (see below). The plots correspond
to the chaotic (p=0.60), critical (p=0.689) and frozen (p=0.8) regimes.

Now in order to characterize the critical properties of our system, we
will applie the well known Derrida's method, based on the annealed
 approximation [15]. This approach is a way to avoid the dynamical
correlations which appear as the system evolves in time. In
the thermodynamic limit the original (quenched) model and the annealed 
counterpart share the same phase transition curves [16].

We start with two randomly chosen configurations
$$C_1(t) \equiv (\sigma_1^{(1)}(t),...,\sigma_N^{(1)}(t))\eqno (8.a)$$
$$C_2(t) \equiv (\sigma_1^{(2)}(t),...,\sigma_N^{(2)}(t))\eqno (8.b)$$
which are also randomly taken from the set ${\cal C}_S(N)$ of all the
possible $N$-strings (clearly $\sharp {\cal C}_S(N) = S^N$).
Following Derrida's method, it can be shown that the overlap
$a_{12}(t)\in [0,1]$, defined as the normalized number 
$N a_{12}(t)$ of elements with common states in $C_1(t)$ and $C_2(t)$, 
will evolve in time following a nonlinear
one-dimensional map. In short, let $N a_{12}(t+1)$ the net overlap after
one iteration of $C_1(t)$ and $C_2(t)$ under (7). Then a new set of
connections and  $\Lambda_i^s $ functions is again taken from
${\cal S_K}$, and a new iteration is performed over
$C_1(t+1)$ and $C_2(t+1)$.

For our $S$-state random network and assuming that $1 \over S$ is the probability
 that two outputs of an arbitrary function $\Lambda_i^s $ are identical, following
 the arguments in [15], we have a nonlinear map for $a_{12}$ given by:
$$a_{12}(t+1) = a_{12}^{K}(t) +
{1 \over S} \Biggl [ 1 - a_{12}^{K}(t)\Biggr ]\eqno (9)$$
where: $a_{12}^K (t)$ is the fraction of elements with identical inputs and
 $ [ 1 - a_{12}^{K}(t)]/S$ is the probability for two automata, with at least
 one different input,  being equal at $t+1$.
Now the critical point where the phase transition takes place (separating
the so called frozen and chaotic phases) is obtained from the stability condition:
$$ \left. {\partial a_{12}(t+1) \over \partial a_{12}(t) } 
\right\vert_{a^*=1}=K \Biggl (1 - {1 \over S} \Biggr ) \leq 1 \eqno (10)$$
or in other words the following relation between the average
network connectivity and the number of available states is reached:
$$ K = {1 \over 1-{1\over S}} \eqno(11)$$
We see that for $S=2$ (Boolean net) we recover the well known critical point
 $K_c=2$. For $S\gg 2$ we can approximate the marginal stability relation to:
$$ K \approx 1+{1 \over S} \eqno (12)$$
We also see that
when $S \rightarrow \infty$ the average connectivity moves to 
$K_c \rightarrow 1$ (i.e. the critical point is reached only at
the lowest connectivity). The previous inequality (10) is
 related with the high-temperature value of damage spreading [23] of the Potts
 model [13].

In RNS it is possible to extend Derrida's $p$-parameter definition:
$p$ is defined as the probability of having a $0$ as output of the function
$\Lambda_i^s$ ($0$ is the quiescent state) and the other states are equally
likely to be present. Although this choice seems too limited, it can be justified 
from variational arguments (see Appendix I). Now the overlap will evolve 
following the 1D map:
$$ a_{12}(t+1) = 
a_{12}^K(t) + \Biggl [ p^2 + {(1-p)^2 \over S-1} \Biggr ] 
( 1 - a_{12}^K (t) ) \eqno(13)$$
where we have:

\begin{itemize}

\item $a_{12}^K (t)$ is the fraction of elements with
 identical inputs at time $t$

\item $( 1 - a_{12}^K (t) )p^2$ is the probability  of two automata with at
 least one different input, 
 being equal to "0" at $t+1$.

\item $( 1 - a_{12}^K (t) )(1-p)^2/(S-1)$ gives the probability of finding two
 automata with at least one
 different input,  being equal (but differents from $0$)
 at $t+1$.

\end{itemize}

Now, the stability analysis for $a_{12}^*=1$ gives us a
 general condition for the critical 
points:
$$ p_c= {1 \over S} \Biggl [ 1 + \Bigl ( 1 - {S \over K}
\Bigl [ (2-S) K + S - 1 \Bigr ] \Bigr )^{1/2} \Biggr ] \eqno(14)$$
Defining a critical surface $p_c=p_c(K,S)$ separating the frozen
(ordered) phase from the chaotic one.

Many possible characterizations of the two phases and the critical
surface can be used. In figure (6a-c) we show three examples of the magnetization $m$ (for different RNS  with $N=100, K=2$ and $S=5$ states). 
Here $m$ is defined as the
number of times $n$ a given automaton $S_i$ is such that $S_i=0$
 over $T$ steps i. e.
 $m=n/T$. Then $P(m)$ is a histogram giving the distribution of
 $m$ for the whole
 system i. e. it counts how many automata with the same average
 magnetization $m$. 
We can see, as expected, how $P(m)$ moves from a singular distribution with few
 peaks to a more continuous distribution at the chaotic regime.

A numerical characterization fo these three regimes can be easily obtained
by computing the frequency distribution $N(T)$ of cycles of different length $T$.
Since these dynamical systems are finite, with a maximum number of $S^N$ states,
the previous equations (7) lead to two types of attractors (cycles
 and steady states). However, the presence of a critical boundary separating the two
 regimes is made clear by the existence of well-defined scaling laws.
 In figure (6d-f)
we show the frequency distribution of periods for the three regimes.
At the chaotic phase, long-periodic orbits are rather common. 
This is due to the exponential increase of attractor lengths in this
phase. The frozen regime shows the opposite tendency: the periods are very
short, with an exponential decay with $T$. At the critical boundary, both 
short and long periods are present, in such a way that in spite of the 
frequent observation of short-period orbits, long-periodic orbits of length
$T \approx O(N)$ are allways found.

In random Boolean networks, the previous observations have been deeply explored
through both numerical and theoretical approximations [6,7]. A specially
important relation is the dependence of the average period $<T>$ with the 
system size $N$. When RBN are used with $p=0.5$, it is found that the critical phase ($K=2$)
is characterized by a scaling $<T> \approx \sqrt{N}$ and an exponential
growth $<T(K>2)> \approx 2^{BN}$ is obtained at the chaotic phase, with $B>1$ [2].
We have explored this scaling behavior in the RNS counterpart. Two examples of
our calculations are shown in figure 8(a-b). The average period length has been
calculated in two different cases, both varying $p$ (a) and the number of states $S$ (b).
we can see how the scaling $<T(N)>\approx N^{\tau}$ is present at criticality
and how an exponential increase in $<T>$ also appears as we enter into the 
chaotic phase. Numerical simulations show that for the chaotic phase an exponential
law $<T(S,N)>\approx S^{BN}$ is obtained, with $B>1$. Numerical estimations of
this value gave $B = 1.11 \pm 0.02$.

\section{CA and RNS}
It can easily be checked from (14) that for $S=2$ (Boolean net) we have
$$ p_c= {1 \over 2}+ {1 \over 2}\sqrt{( 1 - {2 \over K})} \eqno(15)$$
which gives us the well known critical line in the $(K,p)$-phase space [15,17].

Equation (15) is reminiscent of those found in standard 
CA (with nearest-neighbor connectivity [18]). The existence of a transition domain for CA models
has been characterized by means of several approaches. Langton's $\lambda$-parameter gives a rough characterization of these critical points when complex CA rules
(involving high $S$ and/or $r$ values) are used. If $S^{2r+1}$ is the total number of $(2r+1)$-configurations in the CA rule table, as described by (1), and $M$ out of the total map to a non-zero state, then $\lambda$ is defined as [18]:
$$ \lambda \equiv {M \over S^{2r+1}} \eqno(16)$$
Thus, $1-\lambda$ corresponds to $p$ under the RNS definition.
This parameter has been shown to be related with -but not equivalent to-
 a temperature in statistical physics [18]. It was observed (particularly
 in the large-$r$ limit) that these
CA exhibit some of the characteristics of second-order
phase transitions (for a detailed study see [18], although the association
 between computation, complexity and critical phenomena has been shown
 to be flawed [19]).
Li et al. [18] showed by means of mean field approximations that, for
$S=2$, there is a simple relationship between the 
critical value $\lambda_c$ and $r$ (the neighborhood radius):
$$1 - \lambda_c = {1 \over 2} +
{1 \over 2} \sqrt{1 - {2 \over r+1}} \eqno(17)$$
This result is consistent with equation (15). But in (17), instead of
$2r+1$ (the total connectivity of the CA model) we have $r+1$ (half of it
 plus one).
In the RNS counterpart, we have $K$, the total connectivity. These
differences are due to the spatial correlations intrinsic
to CA models, which are destroyed in the RNS counterpart. This is consistent
 with the analysis of Li et at. [18] which is based in a mean-field
 estimation of the spreading rate of diferent CA patterns. Assuming 
 symmetry in the average spreading rate $\gamma(r)$ in right and left
 directions, they obtain the onset to non-zero spreading at $\lambda_c$
 given by (17). The corresponding critical curves for both the $1-\lambda$
 parameter (squares) and $p$ (circles) as a function of
 the connectivity are shown in figure (8). 
Remember that a neighbor radius $r$ in CA is equivalent to
 $K=2r+1$ in RNS. Thus the first point represented for RNS whit $S=2$ is $K=3$ with $p_c=0.79$. Both curves share the
 same qualitative behavior, but larger values of the bias are required
 in RNS in relation with CA in order to reach the ordered regime, 
as expected. 
A similar relationship is found for the
 temperature in relation with $q$ in some $q$-Potts models [13]. 

\section{Stable core and the self-overlap method}

A well-defined critical phase transition should be characterized by an appropiate order
 parameter $\Omega$ such that $\Omega > 0$ at the disordered regime and zero otherwise.
In fact, $d^*=1-a^*$ as like parameter of order for RBN and RNS [15].
 In this context,
Flybvjerg [22] explored a different approach than Derrida's by 
defining the {\em stable core} at time $t$, $c(t)$, as the (normalized)
 fraction of automata $S_i$ (independent of the initial condition) reach stable
 values at a given step $t$, i.e. remain constant after $t$.
The stable core asymptotic behavior, $c(t \rightarrow \infty)$, can be used  as
 an order parameter for the frozen-chaos transition in RBN.
Explicitly, if $\Omega \equiv 1 - c(t \rightarrow \infty)$, the previous requirements
 for the order parameter definition hold.

The argument of Flybvjerg for finding an iterated equation for $c(t)$ is as follows. 
There are $K+1$ mutually exclusive reasons for a given automata $S_i$ to be part
 of the stable core from step $t$ to $t+1$. The probabilities of this $K+1$ reasons are:
$(0)$ All the inputs, 
$\sigma_{i_1} (t), \sigma_{i_2}(t),...,\sigma_{i_K}(t)$, belong to the stable
 core at $t$. Given that the inputs are choosen at random, this situation occur 
with probability $c^K(t)$ (where $c(t)$ is interpreted as the probability of
 belonging to the stable core  in $t$).

$(1)$ $\Lambda_i$ is independent of one input and the rest of the $K-1$
 variables belong to the stable core:
$${K\choose 1} c^{K-1}(t) (1-c(t)) p_1 \eqno(18)$$
where $p_1$ is defined as the probability of $\Lambda_i$ to be
 independent of one input after fixing its $K-1$ remainder inputs.

$\vdots$

$(j)$  $\Lambda_i$ is independent of $j$ inputs and the rest of the $K-j$ variables belong to the stable core:
$${K\choose j} c^{K-j}(t) (1-c(t))^j p_j \eqno(19)$$
where $p_j$ is defined as the probability of $\Lambda_i$ to be independent of $j$ arguments after fixing its $K-j$ remainder arguments.

$\vdots$

$(K)$ $\Lambda_i^s$ is a constant fuction (independent of its $K$ arguments), with all out the stable core: $$(1-c(t))^K p_K \eqno(20)$$
Adding the $K+1$ exclusive reasons:
$$ s(t+1) = \sum_{i=0}^{K} {K\choose i} c^{K-i}(t) (1-c(t))^K p_i \eqno(21)$$
where $p_i$ is the probability that the Boolean function is independent of a 
certain number $i$ of inputs after fixing its $K-i$ remainder inputs
 (a biologically relevant case are the canalizing
functions that depend on a unique input and are independent of remainder [2]).
For the Boolean functions with bias $p$, i.e., with value $0$ with probability 
$p$ and value 1 with probability $(1-p)$ it gives:
$$ p_i = p^{2^i} + (1-p)^{2^i} \eqno(22) $$
By substituying (22) into (21) and analyzing its stability the following 
transition critical curve is obtained: $K 2p(1-p) = 1$
according to [15,23], reached by different methods.

How to extend this argument to RNS? Apparently, such an extension is straightforward.
 It seems that, again, there are $K+1$ exclusive reasons for a given automata to move
 from outside the stable core to be a member of it. Such a simple translation of the
 previous procedure is not possible, and we can show why by means of a simple example.
Let us consider a RNS with $K=4$ and $S=3$, i.e.:
$$\Lambda_i^3=\Lambda_i(\sigma_{i_1}, \sigma_{i_2},\sigma_{i_3}, \sigma_{i_4}) \; \;
 \; ; \; \; \; i=1,2,...,N \eqno(23)$$
and let us assume that at a given $t$ the first three inputs belong to the stable core
, i.e:
 $\sigma_{i_1}, \sigma_{i_2},\sigma_{i_3} \in s(t)$ and 
$\sigma_{i_4} \notin s(t)$. The question is whether or not $\sigma_i (t+1) \in s(t+1)$
 if
 $\sigma_i (t) \notin s(t)$. This will happen if $\Lambda_i^3$ is independent of
 the fourth
 input. But it can also occur (and this makes a big difference) if $\sigma_{i_4}$ can only
 reach a subset $\Sigma' \subset \Sigma$, say $\Sigma' \equiv \{ 0, 1 \}$. In such case,
 we have a reduction in the number of degrees of freedom for $\sigma_{i_4}$. In this case
 it can happen that $\Lambda_i^3$ is independent of $\sigma_{i_4}$ for $\sigma_{i_4} \in
 \Sigma'$ but not if $\sigma_{i_4}=2$ (for $S=2$ a reduction of one degree of freedom
 simply means that the input belongs to the stable core). These possible outcomes forces
 us to weight the probabilities that the input units have their degrees of freedom
 reduced. 
 More generally, the exclusive reason (1) implies that $\Lambda_i^s$ is independent of a
 given input $\sigma_{i_j}$. This occurs with probability  $p_1=p^2+(1-p)^2$ for the case
 of two states. For $S$-states, it can occur that $\sigma_{i_j} \in \Sigma' \subset
 \Sigma$ and $\sigma_{i_j} \notin s(t)$. In order to calculate the order parameter
 equation for networks with multiple states, we will compute the successive overlap
 (or {\em self-overlap} [25]) $a_{t+1,t}$, defined as the fraction of automata such that:
 $\sigma_i(t+1)=\sigma_i(t)$.
This is in fact the overlap between the system configuration at $t$ with the next
 configuration at $t+1$. If the assymptotic value of the stable core is $1$ the
 self-overlap
 $a_{t+1,t}(t\to \infty)$ will be one too. This allows us to use the asymptotic
value of the self-overlap as an appropiate order parameter and as an alternative
to the standard overlap equation analysed in section 2.

The iterated map for the self-overlap will be
$$ a_{t+1,t} = a_{t,t-1}^K + \Biggl ( p^2 + {(1-p)^2 \over S-1} \Biggr)
 (1 - a^K_{t,t-1})  \eqno (24)$$
where we desribe the self-overlap between the set at $t+1$ and $t$ as a function
of the self-overlap for $(t-1,t)$. If we interpret $a_{t,t-1}$ as the
probability for an arbitrary unit to remain in the same state at both $t-1$ and
$t$, the term $a_{t,t-1}^K$ gives the probability that all the inputs of a given
unit are the same from $t-1$ to $t$. Obviously $(1-a^K_{t,t-1})$ is the probability
that at least one of the inputs will be different between $t$ and $t-1$. In that case,
there is still a possibility of remaining in the same state at $t$ and at $t+1$
just by chance. This is given by $(p^2 + (1-p)^2)/(S-1)$. 

The stability analysis now gives:
$$ \left. {\partial a_{t+1,t} \over \partial a_{t,t-1} } 
\right\vert_{a^*}= K
\Biggl [ 1- \Biggl (p^2 + {(1-p)^2 \over S-1}\Biggr) \Biggr ] \leq 1 \eqno(25)$$
which is easily interpreted from a perturbative point of view [23]:
$p^2 + {(1-p)^2 \over S-1}$ is the probability of no-propagation for the minimal
 perturbation (one change of state $l$ to $m$ in one input). Then,
 $K\Bigl [ 1- \Bigl (p^2 + {(1-p)^2 \over S-1}\Bigr) \Bigr ]$
is the mean number of changes generated by the perturbation in one time step.
This interpretation allows to define a Lyapunov exponent for this system (see
Appendix II).

In figures (9-10) we show (continous lines) the asymptotic behavior of both the overlap and 
the self-overlap for different $p$ and $K$ and $S=10$ from equation (24). 
As we can see, the dynamical equation for the self-overlap (24) is the same as
the standard overlap equation (13), 
although it is conceptually different from it. The self-overlap, as it 
occurs with the stable core, can only be applied to the quenched system and
fails with the annealed case. Besides, 
the self-overlap is computationally more efficient than the standard
overlap, which requires the parallel running of the two replicas of the
system.

In figure (9) we have followed the time evolution (black symbols) of overlaps 
between two RNS quenched replicas of size $N=1000$ as in [15]. Each point is compute averaging
$100$ experiments. The connectivity $K=2$
and the number of avalible states $S=10$ are fixed, and the bias $p$ change. The white 
symbols represent identical averages for self-overlap, except for the crosses 
where $N=10.000$ for critical $p=0.70$ in order to show the finite size effect of the 
critical transition. This effect is more acused for the self-overlap as for the overlap.

In figure (12) we show the time evolution, this time fixint the states $S=10$ and the
bias $p=0.81$, of the overlap and self-overlap. Again the continous lines are the theoretical
evolution from equation (29). Black and white symbols correspond to overlap and 
self-overlap respectivily. In this case the variation of the connectivity despite
the transition at $K=3$. Again the crosses ($N=10.000$ for self-overlap) show the 
finite size efecfct.

All our simulations show a very good agreement whit the theoretical evolution of the
overlap (eq.13) and the self-overlap (eq.24). Thus, any of them acts as good order 
parameter.

\section{Summary}

In this paper we have proposed a simple, straighforward extension of random Boolean
 networks in terms of a discrete model with multiple states. The starting point of
 our analysis is the observation that several natural systems involving a wired set
 of entities of some kind often display a range of state values instead of simple on
-off characteristics. In some cases, as the specific tasks (states) performed by ants
 in a colony, the discreteness of the states is rather well defined, although their
 transitions (as a consequence of interactions of ants engaged in different tasks)
 can be rather complex [24] suggesting that the underlying task-dependent rules are
 complicated ones. When cell types are considered, we also get a well-defined set of
 "states" although each one itself is the result of a complex dynamical system (the
 gene expression pattern) at a given stationary state. A classical set of experiments
 with different organisms at different levels of development reported the 
existence of complex interactions between cell types eventually leading to switches
 by depending of the cell types under interaction [2,12]. In other systems, like the
 genomic regulatory network, the activity of some genes is 
sometimes close to an on-off switch
 but it displays a range of activity levels. A first approximation to such continuous
 networks is the RNS described and analysed in this paper. Following previous studies,
 our interest was focused on to the critical features of these networks. It has been
 conjectured that a wide range of complex systems, from evolving ecosystems, 
biological regulation networks, traffic or ant colonies [4], to cite only a few, display
 patterns in space and time suggesting that they are close to phase transition points.
 In this context, we have found the explicit form of the transition domains as well
 as an order parameter equation for the transition.  

The critical boundaries for this model have been obtained from Derrida's annealed
 approximation and the computation of $s(t)$ (the stable core equation) required the
 development of an alternative treatment to the Flyvbjerg derivations for standard RBN
 [22]. This model has been shown to share some common traits with other related dynamic
 models like cellular automata and the $q$-Potts model.

Several extensions of this work are possible. Further quantities can be defined in
 order to characterize the damage spreading. This can be easily done by means of 
theoretical extensions of previous definitions of Lyapunov exponents for RBNs [25].
 One of them is the analysis of the square-lattice counterpart of the RNS with
 nearest-neighbor connectivity. This type of network has been proved to be extremely
 useful in order to display the critical features of the RBN model in terms of 
 percolation on a square lattice [15]. A different extension would be a re-definition
 of the network rules in order to make the transitions from different states closer
 to the continuous dynamics. If the output of the $\Lambda_i^s$ function is randomly
 chosen from $\Sigma$, we must expect  to observe a time evolution of individual units 
in terms of jumps from a given state to any other state. But if we want to retain the
 relationship between the $S$-state model with the corresponding continuous 
counterpart (as described by sets of nonlinear differential equations) a consistent
 increase or decrease in individual activity should be observable. This problem can
 be easily solved through the introduction of a new set of rules describing the outputs
 of the $\Lambda_i^s$-function in terms of increases/decreases of gene activity and
 will be reported elsewhere. 

\vspace{1 cm}

{\large Acknowledgments}

\vspace{0.5 cm}

The authors would like to thank Andy Wuensche, Brian Goodwin, 
Jordi Delgado and Mar Cabeza for useful discussions and help. This work
 has been partially supported by a 
grant DGYCIT PB-97-0693 and by the Santa Fe Institute (RVS and SK) and by
the Centro de Astrobiolog\'{\i}a (BL).

\section{Appendix I: RNS and maximum entropy}

The relation (11) can be easily generalized: $K$ can be not unique, but an
 average connectivity $<K>$ is always able to be defined [17] and we can have a 
probability distribution of states,
 i.e. $P(\mu)=P[\Lambda_i^S=\mu]$ for $\mu \in \Sigma$. In this general case, the
 critical line transition is given by: $<K> = 1/(1 - {\sum {P^{2}(\mu)}})$ 
[23]. This
 result can be used in order to find the expected probability distribution of states
 at criticality [17]. Using the following maximum entropy constraints:
$$\sum_{\mu} {P(\mu)} =1 \eqno (1)$$
$$\sum_{\mu} {P^2(\mu)}= 1 - {1 \over {<K>}} \eqno (2)$$
$P^*_c(\mu)$ can be derived from a variational procedure known as the maximum
 entropy formalism [20]. We first construct the Lagrangian:
$$ {\cal {L}}({\bf P})= - {\sum_{\mu} {P(\mu) \log P(\mu)}} - \beta \biggl
 ({{\sum_{\mu} {P^2(\mu)}} - {1 \over <K>}}\biggr ) - \alpha \biggl
 ( {\sum_{\mu} {P(\mu)}} - 1 \biggr ) \eqno (3)$$
where ${\bf P} = (P(0), P(1), \dots , P(S-1))$ and the variation
$$  {\partial {{\cal {L}}({\bf P})} \over \partial {{\bf P}} } 
= 0 \eqno (4)$$
is performed. This leads to a set of equations
$$\log {P(\mu)} + 2 P(\mu) \beta = -(1+\alpha) \eqno (5)$$
which must be solved for $P(\mu)$. Let us introduce $P(0)=w$
 and $P(\mu \ne 0)= \alpha_\mu w$ with $\mu = 1, 2, \dots , S-1$ with $w \in [0,1]$ and 
$\{ \alpha_\mu \}$ is a set of positive constants such that
$$w (1+ \sum_{\mu = 1}^{S-1} {\alpha_\mu}) = 1 \eqno (6)$$
Then we get:
$$\log w + 2 \beta w = - (1 + \alpha)$$
$$\log w + \log {\alpha_\mu} + 2 \beta {\alpha_\mu} w = -(1+\alpha) \; \; \; \; \; \mu = 1,2, \dots , S-1 \eqno (7)$$
So $\alpha_{\mu} = C$ (constant) $\forall \mu$ and as a consequence
$$ P(0) = w \eqno(8.a)$$
$$P(\mu)={1-w \over S-1} \; \; \; \; \mu=1,2,\dots ,S-1 \eqno (8.b)$$
The most likely probability distribution close to criticality will be a
 delta-shaped $P(\mu)$ with a given state present with probability $w$
 and the rest equally distributed. We should note that this property
 (i.e. a sharply peaked distribution with one or a few predominant states)
 has been reported for the onset of chaos in low-dimensional chaotic
 dynamical systems [21]. Here the probability distributions
 maximize a previously defined complexity measure $\cal C$ and numerical
 simulations of RNS indeed show that such singular distribution with one
 or a few peaks is characteristic at criticality.

\section{Appendix II: Lyapunov exponents for RNS}

If, in the distance method or Derrida's aproximation, we interpret the replica of the one system as a perturbation on the original system [25], we can define a expansion rate of the perturbation in a time $t$ for RNS easilly:
$$ \eta(t) = {d(t+1) \over d(t)} \eqno (1)$$
In a natural way, we can define now the Lyapunov exponent as the temporal average of the logarithm of the expansion rate:
$$ \lambda (T) = {1\over T} \sum_{t=1}^T \log \eta (t) =\log{\sqrt[T] {\prod_{t=1}^T {\eta (t)}}}=\log{\bar{\eta}} \eqno (2)$$ 
If we use the equation for the distance between configurations:
$$d_{12}(t+1)=\Biggl (p^2 + {(1-p)^2 \over S-1}\Biggr)[1-(1-d_{12}(t))^K] \eqno (3)$$
Then the expansion rate in the time $t$ of the perturbation, is:
$$ \eta(t) = { d_{12}(t+1) \over  d_{12}(t)} = {\biggl (p^2 + {(1-p)^2 \over S-1}\biggr ) \{1-[1-d(t)]^K \} \over d(t)} \eqno (4)$$
and approximating for small $d(t)$:
$${(1-d(t))^K} \approx {1-Kd(t)} \eqno (5)$$
we have:
$${\eta(t)} \approx {\Biggl (p^2 + {(1-p)^2 \over S-1}\Biggr)K} \eqno (6)$$
an average constant expansion rate that give us that Lyapunov's exponent:
$$ \lambda = \log {\Biggl [ \Biggl (p^2 + {(1-p)^2 \over S-1}\Biggr)K \Biggr]} \eqno(7)$$
that determines the two classics regimes: $\lambda <0$ (order) and
$\lambda > 0$ (chaos), whit the marginal case $\lambda = 0$, in total concordance with the transition surface. 

There are diferent methods to compute the largest Lyapunov exponent. In order to show consistence we will demonstrate that it is possible compute Lyapunov exponents from the self-overlap in agreement with the previous result.
The Wolf's method [26] is used to estimate numerically Lyapunov exponents from time series. In short, the method is described as follows: get two points of the temporal serie, let us say ${\bf X}(t_1)$ and ${\bf X}(t_2)$ and calculate their distance: $\mid {\bf X}(t_2) - {\bf X}(t_1) \mid$. Assume that $\mid {\bf X}(t_2) - {\bf X}(t_1) \mid < \epsilon$, being $\epsilon > 0$ small. Next, compute the distance, after a time $T$, i.e: $\mid {\bf X}(t_2+T) - {\bf X}(t_1+T) \mid$.  This time, usuallyly, is a fraction of the characteristic period or the time required for the autocorrelation function go to zero. Repeating for $n$ pairs of points and averaging, we obtain an estimation of the Lyapunov exponent:
$$\lambda = {1 \over nT} \sum_{t_{2} \neq t_{1}}^{n}{ \log{ { \mid {\bf X}(t_2+T) - {\bf X}(t_1+T) \mid \over \mid {\bf X}(t_2) - {\bf X}(t_1) \mid } } } \eqno (8)$$
For RNS, we can write down an equation for the normalized Hamming distance between succesive time steps in our system, i.e. for the complementary probability of the auto-overlap: $d_{t,t-1}=1-a_{t,t-1}$. Thus the equation for the self-overlap becomes:
$$ d_{t+1,t} = \Biggl[ 1- \Biggl(p^2 + {(1-p)^2 \over S-1}\Biggr)\Biggr][1-(1-d_{t,t-1})^K]  \eqno (9)$$
If we approximate linearly close to the fixed point $d^*=0$. The function becomes:
$$ d_{t+1,t} =  K\Biggl[ 1- \Biggl(p^2 + {(1-p)^2 \over S-1}\Biggr)\Biggr]d_{t,t-1}  \eqno (10)$$
The iterated equation now gives:
$$ d_{t+T,t+T-1} =  \Biggl \{K\Biggl[ 1- \Biggl(p^2 + {(1-p)^2 \over S-1}\Biggr)\Biggr]\Biggr \}^Td_{t,t-1}  \eqno (11)$$
To compute (8) in our terms, and taking $t_2=t_1+1$, i.e.: a natural step in our system,
we have:
$${d_{t+T,t+T-1} \over d_{t,t-1}}=\Biggl \{K\Biggl [ 1- \Biggl (p^2 + {(1-p)^2 \over S-1}\Biggr)\Biggr]\Biggl \}^T\eqno (12)$$
a constant value that, once introduced in the sum (8), determines the  Lyapunov exponent (7) as sugested previously.

\newpage

\section{References}

\noindent
[1] S. Wolfram, Physica D10 (1984) 1; A. Wuensche, {\em The global dynamics of cellular
automata}, SFI studies in the sciences of complexity, Addison-Wesley, 1992; Wuensche
has done a remarkable research on the basins of attraction of both cellular
automata and random Boolean networks; the basic software, named DDLab, is available
in the website: http://www.santafe.edu/wuensch/ddlab.html.

\vspace{0.2 cm}

\noindent
[2] S. A. Kauffman, J. Theor. Biol. 22 (1969) 437
    S. A. Kauffman, J. Theor. Biol. 44 (1974) 167; 
    S. A. Kauffman, Physica D10 (1984) 145; 
    S. A. Kauffman, Physica D42 (1990) 135; 
    S. A. Kauffman , {\em The Origins of Order} Oxford U. Press (Oxford, 1993);R. Somogyi and C. A. Sniegoski, Complexity 4 (1996) 45

\vspace{0.2 cm}

\noindent
[3] D. Kaplan and L. Glass, {\em Understanding Nonlinear Dynamics}, Springer-Verlag (New York 1995)

\vspace{0.2 cm}

\noindent
[4] R.V. Sol\'e, S.C. Manrubia, B. Luque, J. Delgado and J. Bascompte, Complexity 1(4) (1995) 13

\vspace{0.2 cm}

\noindent
[5] R.J. Bagley and L. Glass, J. theor. Biol. 183 (1996) 269

\vspace{0.2 cm}

\noindent
[6] U. Bastolla and G. Parisi, Physica D 98 (1996) 1; 
    U. Bastolla and G. Parisi, J. Phys. A: Math. Gen. 30 (1997) 5613

\vspace{0.2 cm}

\noindent
[7] A. Battacharjya and S. Liang, Physica D95 (1996) 29; 
    A. Bhattacharijya and S. Liang, Phys. Rev. Lett. 77 (1996) 1644

\vspace{0.2 cm}

\noindent
[8] J.A. de Sales, M.L. Martins and D.A. Stariolo Phys. Rev. E 55 (1997) 3262. In this paper the authors study the overall behavior of gene networks by means of a model involving both short- and long-range couplings among genes. They found a rich variety of behaviors and their results consistently fitted known scaling laws for the number of differentiated cells and cell length cycles as a function of $N$. 

\vspace{0.2 cm}

\noindent
[9] K.E. Kurten, Phys. Lett. A 129, 157 (1992) 

\vspace{0.2 cm}

\noindent
[10] S.L. Adhya and D. F. Ward, Prog. Nucleic Acid Res. Mol. Biol. 26 (1981) 103
 
\vspace{0.2 cm}

\noindent
[11] L. Glass and S.A. Kauffman, J. theor. Biol. 39 (1973) 103; see also L. Glass and S. A. Kauffman, J. Theor. Biol. 34 (1972) 219 

\vspace{0.2 cm}

\noindent
[12] S. F. Gilbert, {\em Developmental Biology}, Sinauer, Massachusetts (1996); B.C. Goodwin Temporal organization in cells London, Academis Press (1963)

\vspace{0.2 cm}

\noindent
[13] F. Y. Wu, Rev. Mod. Phys. 54 (1982) 235; C. Tsallis and A. C. N. de Magalhaes, Phys. Rep. 268 (1996) 305

\vspace{0.2 cm}

\noindent
[14] J. E. Lewis and L. Glass, Int. J. Bif. Chaos 1 (1991) 477; T. Mestl, R. J. Bagley and L. Glass, Phys. Rev. Lett. 79 (1997) 653; A. Wagner, Proc. Natl. Acad. Sci. USA 91 (1994) 4387 . These authors consider a simple formulation of biological regulatory networks in terms of a neural-like dynamics, as defined by:
$ {d x_i \over dt} = - x_i \Phi_i \Bigl ( 
\sum_j W_{ij} x_j - \Theta_i \Bigr )$, with $i,j=1,2,...,N$. As usual $\Phi(z)$ is a sigmoidal function and $\theta$ a threshold. The matrix $W_{ij} \in \Re$ describes the specific features of the interactions among different elements (genes). We can see that equations (4.a-b) are just a specific example with $N=2, \Theta_i=0, W_{ij}=1$ and $\Theta(z)=\mu^n/(\mu^n + z^n)$.

\vspace{0.2 cm}

\noindent
[15] B. Derrida and Y. Pomeau, Europhys. Lett. 1 (1986) 45
     B. Derrida and Y. Pomeau, Europhys. Lett. 2 (1986) 739
     G. Weisbuch and D. Stauffer, J. Physique 48 (1987) 11-18
\vspace{0.2 cm}

\noindent
[16] H.J. Hilhorst and M. Nijmajer, J. Physique 48 (1987) 185 

\vspace{0.2 cm}

\noindent
[17] R. V. Sol\'e' and B. Luque, Phys. Lett. A196 (1995) 331

\vspace{0.2 cm}

\noindent
[18] C. G. Langton, Physica D52 (1990) 12
     W. Li, N. H. Packard and C. G. Langton, Physica D45 (1990) 77
     W. K. Wootters and C. G. Langton, Physica D45 (1990) 95

\vspace{0.2 cm}

\noindent
[19] A. Dhar, P. Lakdawala, G. Mandal and S. R. Wadia. Phys. Rev. E51 (1995) 3032

\vspace{0.2 cm}

\noindent
[20] H. Haken, Information and self-organization, Springer, Berlin, 1988; J. N. Kapur, Maximum Entropy models in Science and Engineering, Wiley, New Delhi, 1993

\vspace{0.2 cm}

\noindent
[21] C. Anteneodo and A. R. Plastino, Phys. Lett. A223 (1996) 348

\vspace{0.2 cm}

\noindent
[22] H. Flybjerg, J. Phys. A: Math. Gen. 21 L955 (1988)

\vspace{0.2 cm}

\noindent
[23] B. Luque and R. V. Sol\'e, Phys. Rev. 55, 1 (1997) 

\vspace{0.2 cm}

\noindent
[24] R. V. Sol\'e and O. Miramontes, Physica D80 (1995) 171 and references cited

\vspace{0.2 cm}

\noindent
[25] B. Luque and R. V. Sol\'e, adap-org/9907001  

\noindent
[26] A. Wolf, J. B. Swift, H. L. Swinney and J. A. Vastano, Physica D16 (1985) 285

\vspace{0.2 cm}

\newpage

\begin{figure}[htb]
\caption{Gene network with $N=3$. Here the proteins ($P1,P2,P3$) obtained
from each gene interact in complex ways with DNA. The gene
products bind some specific sites either in the DNA sequence or
on another protein site, thus inhibiting or stimulating gene activity.
Several genes can be involved in the regulation of a given gene (as
described in RBN models through the connectivity $K$. But such
regulation can have different effects on different genes and under
different initial conditions. Here genes 2 and 3 act synergistically
stimulating gene 3 in such a way that the latter produces at a rate 
twice the one expected if gene 2 were repressed by gene 1. If gene
1 starts producing an enough large concentration of $P1$, then it would
be able to supress the stimulation from gene 2 to 3 and thus gene 3
would produce a basal level of protein. If we identify such differences
of gene activity levels as gene {\em states} then we can map the previous
example into a discrete model with a number of (multiple) states}
\end{figure}

\begin{figure}[htb]
\caption{Three-dimensional vector fields for the nonlinear system (6) for $m=3, \mu=1/2$ and
two different values of the nonlinearity $n$: $n=1$ (upper diagram) and $n=4$ (lower). For
a small degree of nonlinearity a single attractor is present, but an increase in the
strenght of the interactions leads to a multiplicity of attractors, both in some of the corners
of the cube and inside it}
\end{figure}

\begin{figure}[htb]
\caption{Space-time diagram for a chaotic RNS, with $N=100, S=5, K=2$
and bias $p=0.60$}
\end{figure}

\begin{figure}[htb]
\caption{Space-time diagram for a RNS at the critical boundary, with $N=100, S=5, K=2$ and bias $p=0.689$}
\end{figure}

\begin{figure}[htb]
\caption{Space-time diagram for a RNS at the frozen regime, with $N=100, S=5, K=2$
and bias $p=0.80$}
\end{figure}

\begin{figure}[htb]
\caption{(a-c) distribution of magnetization $m$ for the three different
regimes of a RNS with $K=2, S=5$. 
Here: (a) $p=0.60$, chaos, (b) $p=0.689$ , critical and (c)
$p=0.8$, frozen regime, respectively. These distributions have been calculated
using a $N=100$ network and averaging over $\tau=10^4$ steps after $10^4$
transients have been removed. In (d-f) the corresponding distributions of
cycle lengths are shown for the same parameter combinations. $M=10^4$ different
RNS have been used, with random connectivities and random initial conditions}
\end{figure}

\begin{figure}[htb]
\caption{Average period length for RNS with (a) $S=5, K=2$ and different biases  
and for (b) $p=0.5, K=2$ and different numbers of states}
\end{figure}

\begin{figure}[htb]
\caption{ Phase diagram obtained from mean field theories for both RNS
 and CA models ($S=2$). Here the fraction of transitions towards
 the quiescent state $0$ is plotted against neighbor radius}
\end{figure}

\begin{figure}[htb]
\caption{Continous lines: dynamical evolution of the self-overlap of RNS of $K=2$
 and $S=10$ with initial self-overlap (and overlap) of $0.1$ and differents values of
 the bias $p$ through the iteration of equation (24).
The black symbols shows numerical averages of the overlap between two RNS quenched 
replicas for $100$ experiments with RNS of size $N=1.000$. The white symbols shows 
identical averages and conditions for self-overlap, except for the crosses 
where $N=10.000$ for critical $p=0.70$ in order to show the finite size effect of the 
critical transition} 
\end{figure}

\begin{figure}[htb]
\caption{Continous lines: dynamical evolution of the self-overlap of RNS of $p=0.81$
 and $S=10$ with initial self-overlap of $0.5$ and differents values of
 the connectivity $K$ through the iteration of equation (24).
The black symbols shows numerical averages of the overlap between two RNS quenched 
replicas for $100$ experiments with RNS of size $N=1.000$. The white symbols shows 
identical averages for self-overlap, except for the crosses where $N=10.000$ for critical
connectivity $K=3$ in order to despite the finite size effect of the critical transition}
\end{figure}

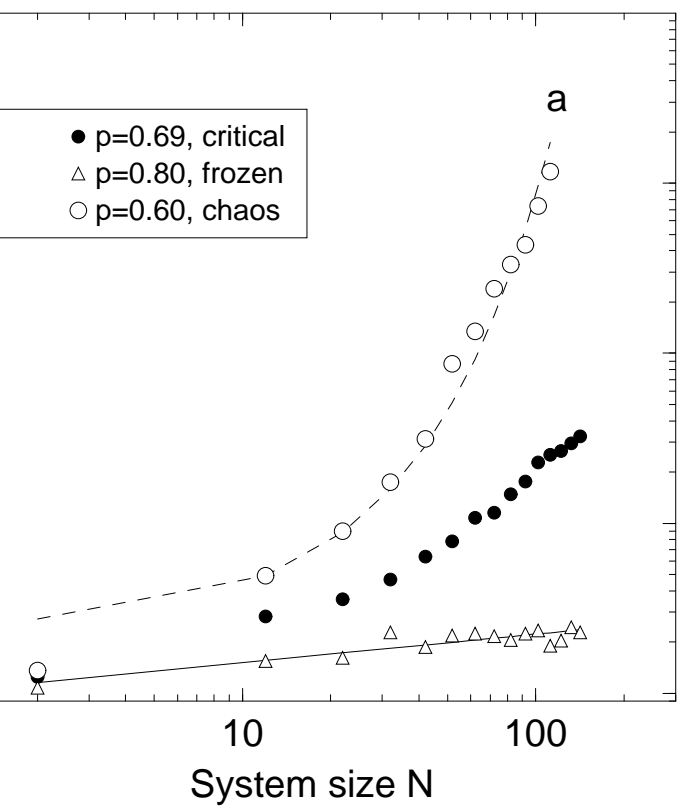
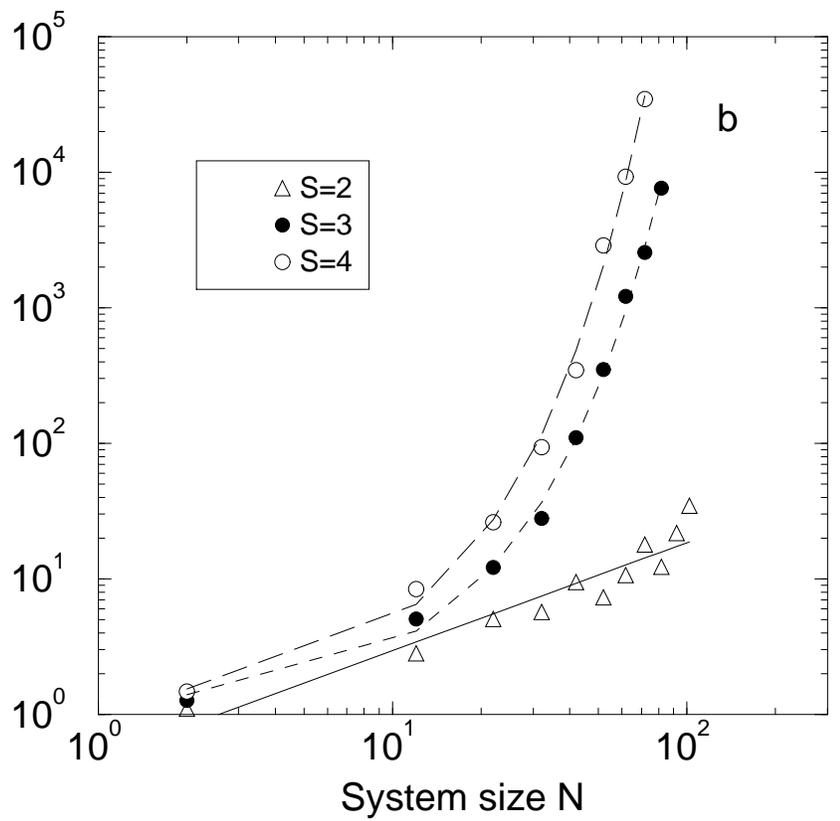
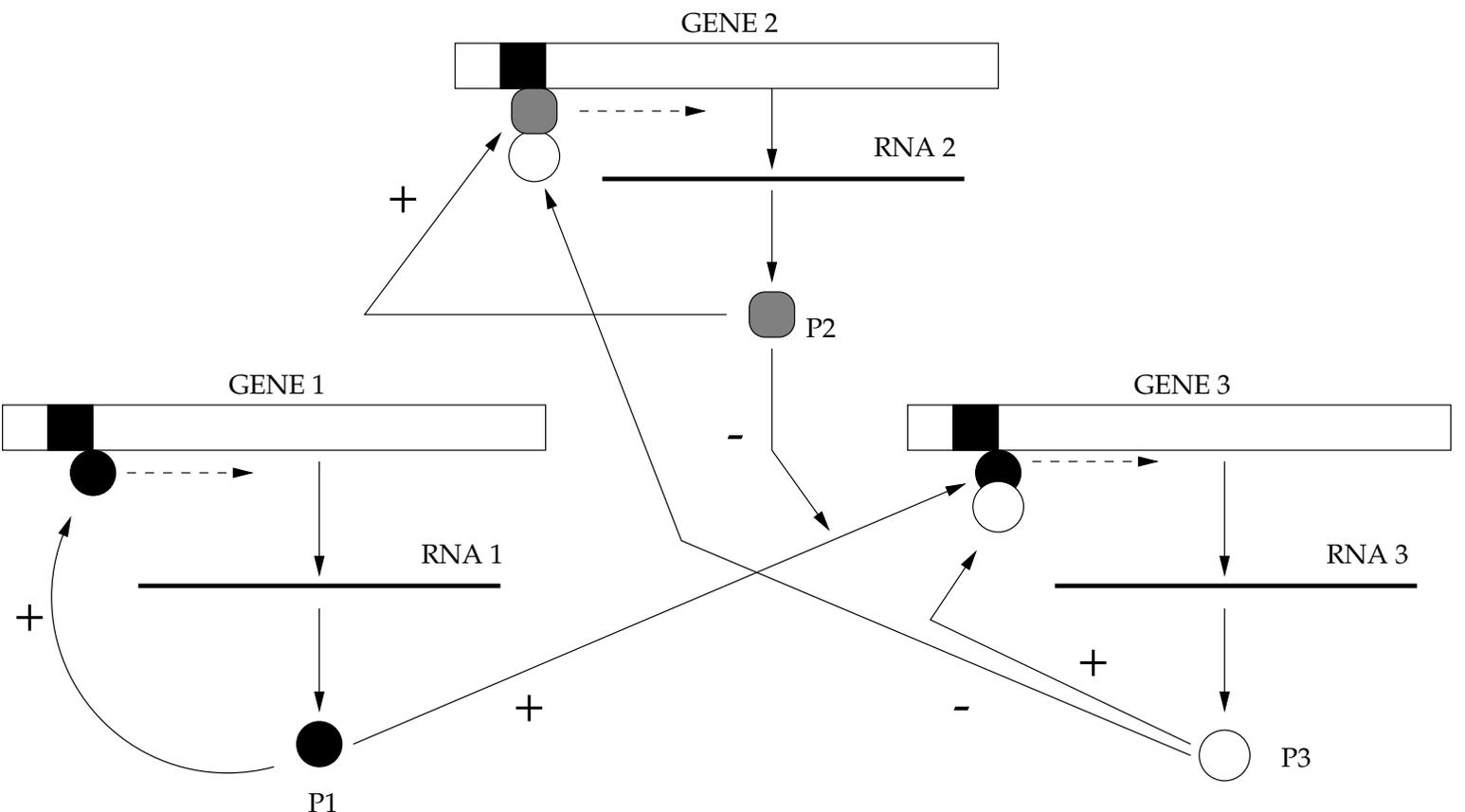
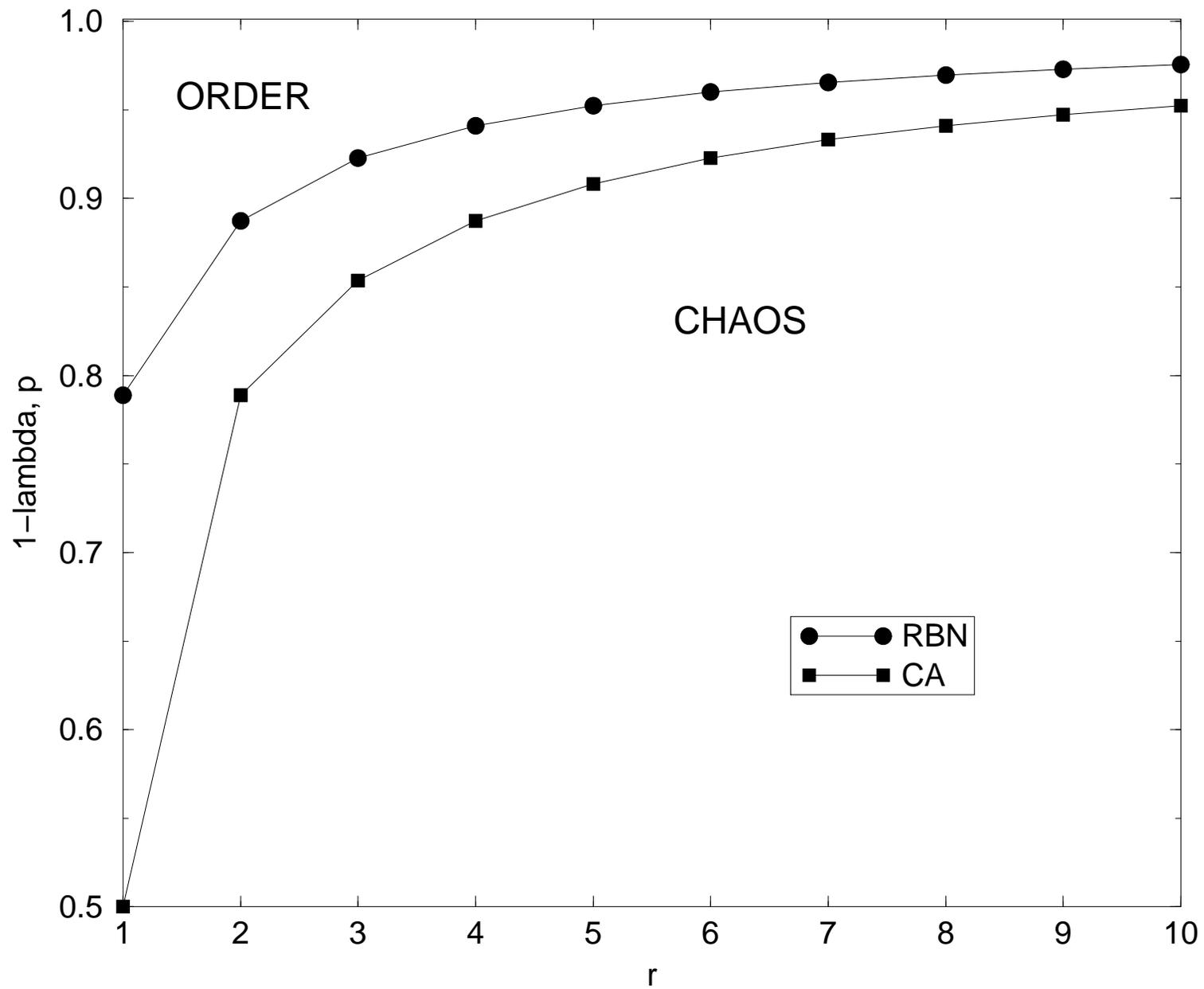
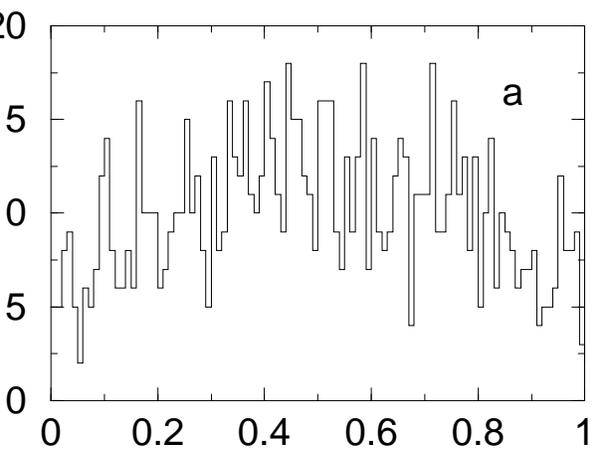
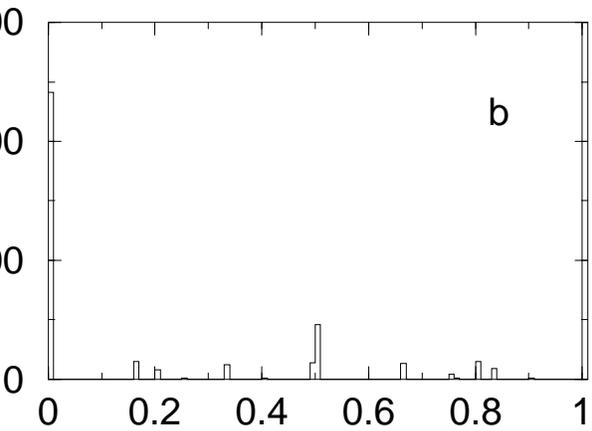
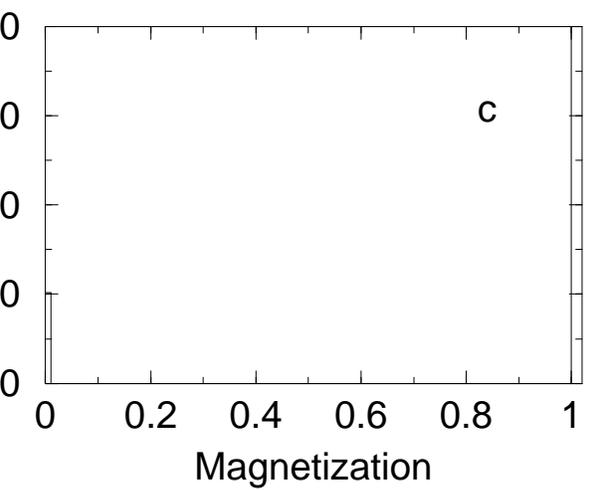
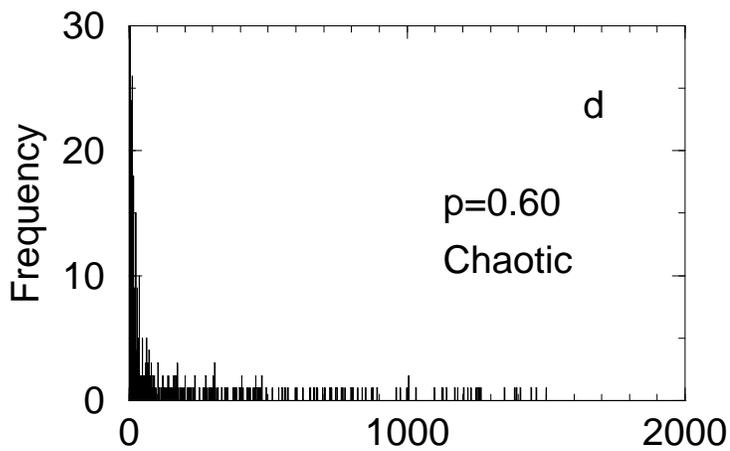
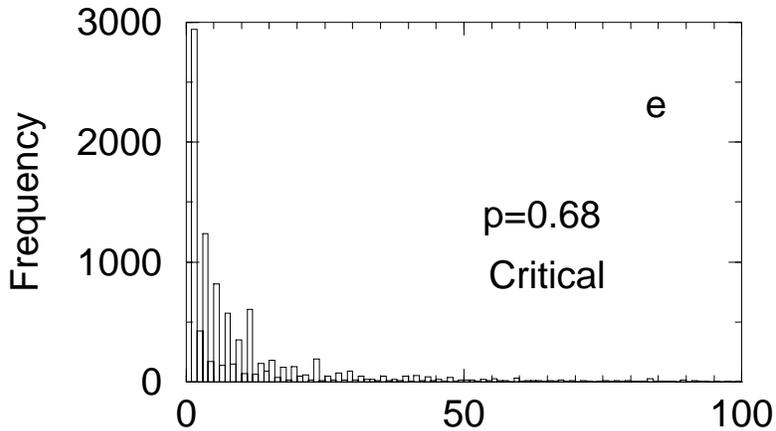
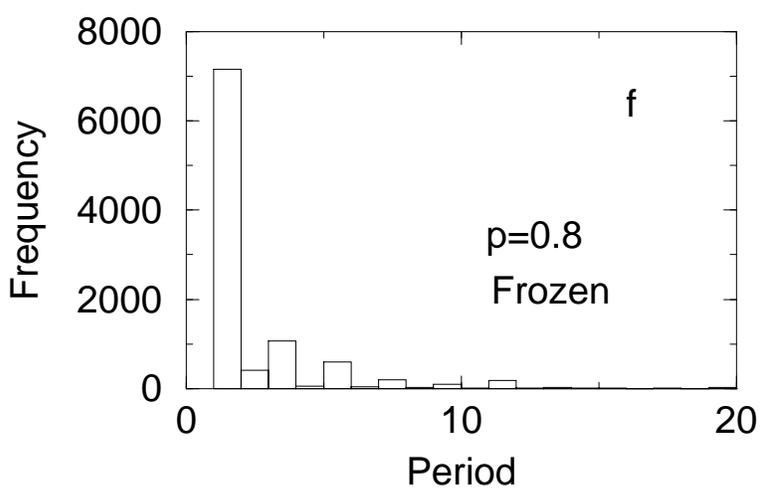
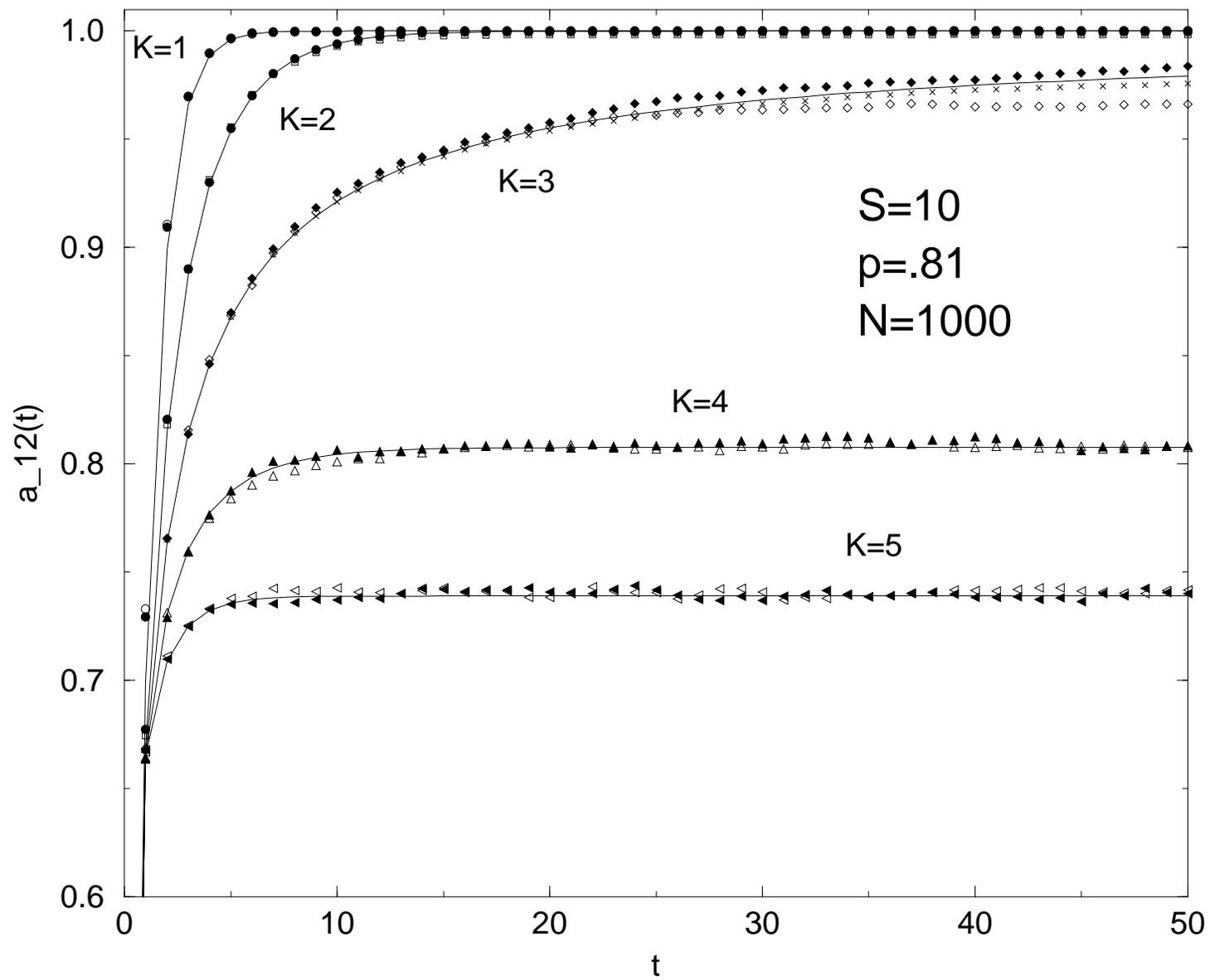
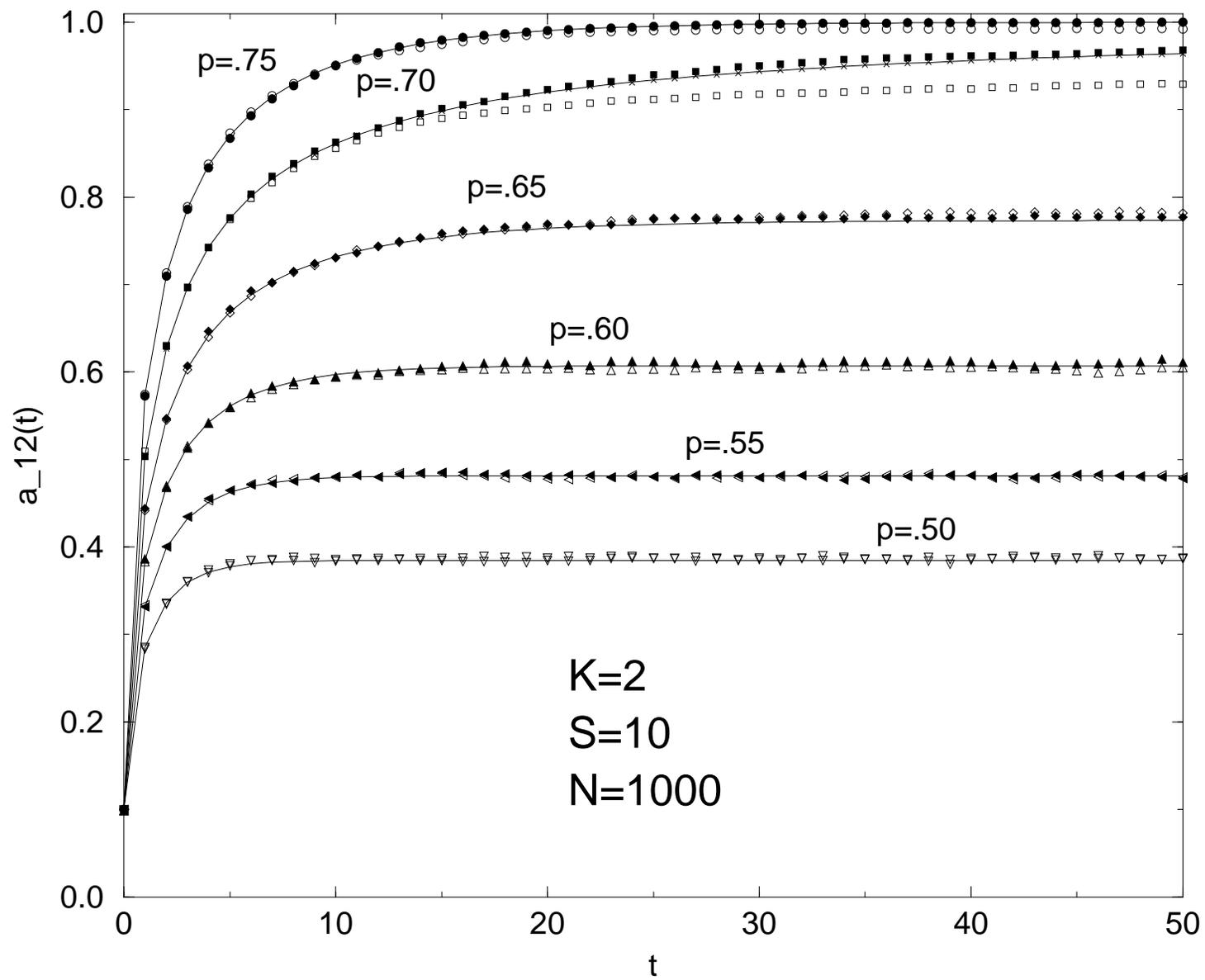
\end{document}